\documentclass[sigconf,screen]{acmart}

\usepackage{multirow}
\usepackage[inline]{enumitem}
\usepackage{listings}
\usepackage{ragged2e}

\newcommand{\heading}[1]{\textbf{#1}.}

\makeatletter
\let\OldTexttt\texttt
\renewcommand{\texttt}[1]{%
  \in@{ }{#1}%
  \ifin@
    \OldTexttt{#1}%
  \else
    \begingroup\ttfamily
    \@tfor\@code@char:=#1\do{%
      \expandafter\texttt@break\@code@char}%
    \endgroup
  \fi}
\newcommand*\texttt@break[1]{%
  \ifnum`#1>64\relax \ifnum`#1<91\relax \discretionary{}{}{}\fi\fi
  #1%
  \ifnum`#1=46\relax \discretionary{}{}{}\fi
  \ifnum`#1=95\relax \discretionary{}{}{}\fi}
\makeatother

\author{Shaojie Jiang}
\orcid{0000-0001-6842-6583}
\affiliation{%
  \institution{University of Amsterdam}
  \city{Amsterdam}
  \country{The Netherlands}
}
\affiliation{%
  \institution{AI Colleagues}
  \city{Diemen}
  \country{The Netherlands}
}
\email{s.jiang@ai-colleagues.com}

\author{Svitlana Vakulenko}
\orcid{0000-0002-5278-8886}
\affiliation{%
  \institution{WU Vienna University of Economics and Business}
  \city{Vienna}
  \country{Austria}
}
\email{svitlana.vakulenko@wu.ac.at}

\author{Maarten de Rijke}
\orcid{0000-0002-1086-0202}
\affiliation{%
  \institution{University of Amsterdam}
  \city{Amsterdam}
  \country{The Netherlands}
}
\email{m.derijke@uva.nl}

\renewcommand{\shortauthors}{Shaojie Jiang, Svitlana Vakulenko, \& Maarten de Rijke}
\copyrightyear{2026}
\acmYear{2026}
\setcopyright{cc}
\setcctype{by}
\acmConference[SIGIR '26]{Proceedings of the 49th International ACM SIGIR Conference on Research and Development in Information Retrieval}{July 20--24, 2026}{Melbourne, VIC, Australia}
\acmBooktitle{Proceedings of the 49th International ACM SIGIR Conference on Research and Development in Information Retrieval (SIGIR '26), July 20--24, 2026, Melbourne, VIC, Australia}

\acmDOI{10.1145/3805712.3808613}
\acmISBN{979-8-4007-2599-9/2026/07}


\title{Orcheo: A Modular Full-Stack Platform for Conversational Search}

\begin{abstract}
Conversational search (CS) requires a complex software engineering pipeline that integrates query reformulation, ranking, and response generation.
CS researchers currently face two barriers: the lack of a unified framework for efficiently sharing contributions with the community, and the difficulty of deploying end-to-end prototypes needed for user evaluation.
We introduce Orcheo, an open-source platform designed to bridge this gap.
Orcheo offers three key advantages:
\begin{enumerate*}[label=(\roman*)]
\item \textbf{A modular architecture} promotes component reuse through single-file node modules, facilitating sharing and reproducibility in CS research;
\item \textbf{Production-ready infrastructure} bridges the prototype-to-system gap via dual execution modes, secure credential management, and execution telemetry, with built-in AI coding support that lowers the learning curve;
\item \textbf{Starter-kit assets} include 45+ off-the-shelf components for query understanding, ranking, and response generation, enabling the rapid bootstrapping of complete CS pipelines.
\end{enumerate*}
We describe the framework architecture and validate Orcheo's utility through case studies that highlight modularity and ease of use.
Orcheo is released as open source under the MIT License at \url{https://github.com/AI-Colleagues/orcheo}.
\end{abstract}

\begin{document}

\maketitle

%!TEX root = ../main.tex

\section{Introduction}
\label{sec:introduction}

Conversational search (CS) has transformed traditional information retrieval (IR) into a dynamic, multi-turn dialogue, allowing users to iteratively refine their information needs~\citep{anand2020conversational,gao2022neural,vakulenko2021mixed}.
The promise of an end-to-end CS system brings significant architectural challenges.
To function effectively, these systems must orchestrate a complex pipeline that includes query rewriting, initial passage retrieval, re-ranking, and dialogue state management~\citep{gao2022neural,vakulenko2019qrfa}.
While research has yielded significant advances in individual components, such as dense retrieval~\citep{karpukhin2020dense}, sparse representations~\citep{formal2021splade}, and contextual query rewriting~\citep{yu2020few,anantha2021qrecc}, the compounded impact of these innovations on the end-to-end CS pipeline remains underexplored.
Consequently, it is often unclear how local optimizations translate to global improvements in user experience.

We identify two persistent challenges that hinder progress, adoption, and impact in CS research:

\heading{Challenge 1: Lack of a unified modular framework}
Advancements in individual CS components are often developed within isolated, ad hoc pipelines, obscuring their true utility and complicating cross-paper comparisons.
This fragmentation forces researchers to reinvent the underlying infrastructure for every new experiment, leading to inconsistent implementations and a ``reproducibility gap.''
There is a critical need for reusable, standardized frameworks that lower the barrier to entry and provide a stable foundation for establishing reliable baselines~\citep{zhang2021chatty,alessio2023decaf}.

\heading{Challenge 2: Barriers to deployment and holistic evaluation}
Moving from a research prototype to a functional application is a major challenge.
It requires engineering skills, such as backend orchestration and web development, that often fall outside the scope of core research.
Without standardized deployment, researchers cannot easily run user-facing studies or apply robust conversational evaluation protocols~\citep{penha2020challenges,faggioli2021hierarchical}.
This lack of empirical feedback creates a bottleneck that prevents theoretical models from being systematically tested and stifles the iterative insights needed to refine future frameworks.

To address these challenges, we introduce \textbf{Orcheo}, an open-source platform designed to unify and accelerate CS research \emph{and} development.
Orcheo treats CS pipelines as first-class graph-structured workflow artifacts that can be easily shared, versioned, and collaboratively improved.
We offer three main contributions:

\begin{enumerate}[leftmargin=*,nosep]
  \item \textbf{A modular CS framework designed for reproducibility.}
Orcheo is a Python platform built on LangGraph that lets researchers package contributions as plug-and-play, single-file modules.
This modular architecture enables seamless integration of new models, such as query rewriters or re-rankers, without the overhead of a monolithic codebase, enabling replication and comparison.
Additionally, Orcheo integrates with an open-source ``Agent Skills'' repository (\url{https://agentskills.io}) to support AI-assisted ``vibe-coding,'' where coding assistants guide users through development and deployment.
  
  \item \textbf{Research-to-production pipeline.} 
Orcheo provides a full-stack environment with dual execution modes: a local CLI for experimentation and a remote backend for production.
To simplify engineering, it includes an encrypted Credential Vault, automated telemetry, and integrated tooling for systematic prompt engineering.
This integrated stack bridges the gap between isolated prototypes and deployable applications.

  \item \textbf{An extensive CS starter kit.}
  We provide a curated library of 45+ off-the-shelf nodes covering the CS lifecycle, from query understanding to response generation, enabling rapid prototyping and benchmarking.
\end{enumerate}

\noindent%
While optimized for CS, Orcheo's domain-agnostic, graph-based architecture can also support adjacent tasks (e.g., conversational recommendation) by swapping in domain-specific nodes.
Orcheo's primary value lies in its broader ecosystem: by unifying algorithms with execution environments, it moves the community beyond ``code sharing'' to ``system sharing,'' ensuring experiments remain executable and extensible after publication.

Section~\ref{sec:related-work} reviews related work. Section~\ref{sec:framework} details Orcheo's node architecture. Section~\ref{sec:platform} describes the full-stack platform. Section~\ref{sec:case-studies} validates Orcheo through case studies, and Section~\ref{sec:conclusion} concludes.

%!TEX root = ../main.tex

\section{Related Work}
\label{sec:related-work}

We review relevant CS research platforms and identify the gaps Orcheo addresses.
Orcheo uniquely integrates modular CS pipelines, full-stack deployment, and built-in IR evaluation metrics, which are typically fragmented across existing frameworks (Table~\ref{tab:comparison}).

\heading{IR Toolkits for Retrieval Modeling}
The IR community relies on mature retrieval toolkits such as Pyserini~\citep{lin2021pyserini} (Python bindings for Anserini~\citep{yang2017anserini}) and PyTerrier~\citep{macdonald2020pyterrier} (a declarative framework for composing complex IR pipelines).
Pyserini is built on Apache Lucene through Anserini, while PyTerrier centers on the Terrier platform with interoperability to other backends.
Both are widely used for reproducible IR research, supporting traditional sparse models (e.g., BM25) and neural retrieval pipelines.

Specialized libraries have helped to advance neural IR: sentence-transformers~\citep{reimers2019sentence} for dense embeddings, and OpenMatch-v2~\citep{yu2023openmatchv2} for training and fine-tuning neural models.
Standardized evaluation and management are supported by BEIR~\citep{thakur2021beir} for zero-shot benchmarks and XpmIR~\citep{zong2023xpmir} for modular neural IR experimentation.

While the established toolkits listed above excel at batch evaluation and static ranking, they are not architected for the multi-turn, interactive nature of CS.
They lack built-in dialogue management for tracking context across turns and the real-time deployment infrastructure required for user-facing applications.
Orcheo addresses these gaps with a unified, graph-based architecture that supports stateful CS workflows and provides extensible interfaces for integrating external retrieval and embedding components.

The effort required to integrate disparate tools creates a \textbf{standardization gap}, not just an engineering burden.
When researchers use ad hoc scripts to connect components, subtle variations in context handling make it difficult to isolate the impact of specific algorithmic contributions.
Orcheo bridges this gap by providing a standardized platform where modular innovations are evaluated within a consistent, reproducible system.

\textbf{RAG Orchestration Frameworks}.
The rapid growth of retrieval-augmented generation (RAG) has introduced orchestration frameworks that link retrieval with generation.
LangChain~\citep{langchain} is widely used for composing LLM-based pipelines, often integrating with LlamaIndex~\citep{llamaindex} for LLM-driven document indexing.
Similarly, Haystack~\citep{haystack} offers a structured, pipeline-based approach for diverse NLP applications.

While powerful, these frameworks prioritize application development over research, creating gaps that Orcheo addresses:
\begin{enumerate*}[label=(\roman*)]
\item \emph{Ease of use}: They often require manual, code-heavy configurations rather than the AI-assisted ``vibe-coding'' and transparency needed for rapid experimentation.
\item \emph{Evaluation infrastructure}: Most orchestrators focus on execution but lack built-in, standardized evaluation routines.
\item \emph{CS-specific nodes}: General RAG frameworks lack native components for specialized conversational tasks like contextual query rewriting that researchers frequently reuse.
\end{enumerate*}

Orcheo builds on LangGraph~\citep{langgraph}, which adds stateful, graph-based orchestration to LangChain and is essential for the iterative nature of CS.
For deployment, LangGraph Server~\citep{langgraph_server} provides a commercial infrastructure for streaming and multi-tenancy.

\heading{Conversational Search and Conversational AI Frameworks}
Several frameworks support conversational AI and CS research.
The closest CS-specific frameworks to Orcheo are Chatty Goose~\citep{zhang2021chatty} and DECAF~\citep{alessio2023decaf}.
Chatty Goose codifies a canonical multi-stage conversational passage retrieval pipeline with modular rewriter, retriever, and reranker interfaces plus reproducible CAsT baselines, while DECAF broadens the experimentation space with a more explicitly modular architecture and both sparse and dense retrieval components.
These systems are important precursors and are the most direct conceptual comparisons to Orcheo.
Orcheo differs in scope by combining CS-specific modularity with a full-stack execution environment, deployment tooling, credential management, and workflow-level evaluation nodes in a single platform.

ConvLab~\citep{lee2019convlab} and ConvLab-2~\citep{zhu2020convlab2} offer modular toolkits for task-oriented dialogue, but focus on slot filling rather than open-domain search.
ParlAI~\citep{miller2017parlai} provides a unified platform for dialogue datasets and agents, though its emphasis is on language models rather than IR.
More general-purpose platforms like Rasa~\citep{bocklisch2017rasa} and DeepPavlov~\citep{burtsev2018deeppavlov} provide production-ready NLU and dialogue management for building conversational assistants, yet they are designed around intent classification and skill-based architectures rather than retrieval pipelines.

These frameworks support dialogue research but lack explicit support for the full CS lifecycle, from query processing and retrieval through re-ranking and grounded response generation.
Orcheo fills this void by providing a unified environment for designing modular, shareable workflows as dialogue-aware graphs.

\heading{IR Reproducibility Platforms}
The IR community has standardized evaluation through platforms like TIREx~\citep{frobe2023tirex} and its infrastructure layer, TIRA.io~\citep{frobe2023tira}, which enable large-scale evaluation via containerized submissions.
These are bolstered by TIREx Tracker~\citep{hagen2025tirextracker} for automatic metadata capture and ranxhub~\citep{bassani2023ranxhub} for sharing and comparing run files.
These platforms excel at archiving results, but they do not address the complexities of designing and packaging reproducible CS workflows.
Orcheo complements them by serving as the environment that produces the research artifacts these platforms are intended to evaluate.

\begin{table*}[t]
\centering
\caption{Comparison of Orcheo with related frameworks and platforms.}
\label{tab:comparison}
\small
\begin{tabular}{p{2.5cm}p{3.5cm}p{2.1cm}p{1.9cm}p{2cm}p{2.3cm}}
\toprule
\textbf{Framework} & \textbf{Primary focus} & \shortstack[l]{\textbf{CS pipeline}\\\textbf{support}} & \shortstack[l]{\textbf{Modular}\\\textbf{sharing}} & \shortstack[l]{\textbf{Deployment}\\\textbf{tools}} & \shortstack[l]{\textbf{Evaluation}\\\textbf{support}} \\
\midrule
\textbf{Orcheo} & Workflow orchestration & \textbf{Native} & \textbf{Node registry} & \textbf{Full-stack} & \textbf{Built-in nodes} \\
DECAF~\citep{alessio2023decaf} & CS experimentation & \textbf{Native} & Components & N/A & CAsT baselines \\
Chatty Goose~\citep{zhang2021chatty} & CS passage retrieval & \textbf{Native} & Components & Via ParlAI & CAsT '19 baselines \\
ConvLab-2~\citep{zhu2020convlab2} & Task-oriented dialogue & Partial & Plugin system & Limited & Dialogue metrics \\
ParlAI~\citep{miller2017parlai} & Conversational AI & Limited & Agent zoo & Limited & Dialogue metrics \\
Pyserini~\citep{lin2021pyserini} & Sparse/dense retrieval & Component only & N/A & N/A & External \\
LangChain~\citep{langchain} & LLM applications & Via composition & Hub & External & External \\
LangGraph Srv.~\citep{langgraph_server} & Graph workflow deployment & Via composition & Via code & \textbf{Server/Cloud} & External \\
LlamaIndex~\citep{llamaindex} & Data + retrieval & Via composition & Hub & External & External \\
Haystack~\citep{haystack} & NLP pipelines & Via composition & Components & Cloud offering & External \\
TIREx~\citep{frobe2023tirex} & IR evaluation & N/A & N/A & Evaluation only & \textbf{Comprehensive} \\
\bottomrule
\end{tabular}
\end{table*}

%!TEX root = ../main.tex

\section{The Orcheo Framework}
\label{sec:framework}

%Orcheo is a LangGraph-based framework that enables researchers to define CS components as modular, reusable \emph{nodes}.
%In this Section, we provide more details on the node architecture, state management, and mechanisms for sharing contributions.

This section details the architectural foundations of Orcheo.
By leveraging LangGraph, Orcheo formalizes CS pipelines as stateful directed graphs in which each component is encapsulated as a discrete, interchangeable node.

\subsection{Design Principles}

Orcheo's framework is built around three core principles directly motivated by the challenges identified in Section~\ref{sec:introduction}:

\begin{enumerate}[label=\textbf{(P\arabic*)}, leftmargin=*]

\item \emph{Minimal Shareable Code}:
Researchers should be able to share novel algorithmic contributions as a single Python file implementing a node.

\item \emph{Composability with Existing Components}:
To foster an ecosystem for reproducible research, new nodes should integrate seamlessly with existing upstream and downstream components.
%A new re-ranker should work with any retrieval node; a new query processor should feed into any downstream retrieval strategy.

\item \emph{Type-safe Interfaces}:
To move beyond ``ad hoc scripts,'' Orcheo enforces strict schema validation.
By explicitly typing node inputs and outputs, we move error detection from runtime to development time, catching integration errors early and enabling automated compatibility tests.

\end{enumerate}

\subsection{Node Architecture}

\begin{figure}[t]
  \centering
  \includegraphics[width=0.98\linewidth]{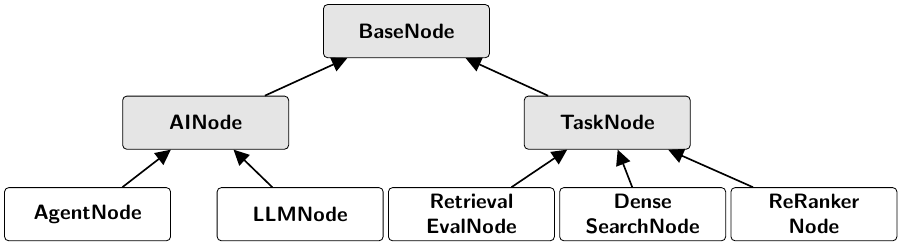}
  \caption{Orcheo's node hierarchy. BaseNode provides shared infrastructure; AINode and TaskNode specialize for LLM-based and utility operations, respectively.}
  \Description{Class hierarchy diagram showing BaseNode at the top, AINode and TaskNode as intermediate abstractions, and example concrete nodes (AgentNode, LLMNode, RetrievalEvaluationNode, DenseSearchNode, ReRankerNode) at the bottom.}
  \label{fig:node-hierarchy}
\end{figure}

Orcheo provides a three-tier node hierarchy designed for different use cases (Figure~\ref{fig:node-hierarchy}):

\heading{BaseNode}
Provides core functionality shared by all nodes:
\begin{enumerate*}[label=(\roman*)]
  \item \emph{Variable interpolation}: Node attributes can reference outputs from previous nodes using \texttt{\{\{node\_name.field\}\}} syntax (e.g., \texttt{\{\{retriever.documents\}\}}), enabling dynamic configuration.
  \item \emph{Credential resolution}: Secure access to API keys and secrets via \texttt{[[name]]} or \texttt{[[name\#field]]} references, resolved at runtime.
  \item \emph{Tool interface}: Nodes can be exposed as tools for agent nodes.
\end{enumerate*}

\heading{AINode}
Extends BaseNode for nodes that involve LLM interactions (e.g., query rewriting, response generation).
Outputs are wrapped in LangChain message format to ensure compatibility with conversation history.
Subclasses implement an async \texttt{run()} method that returns messages.

\heading{TaskNode}
Extends BaseNode for utility and integration nodes (e.g., retrieval, re-ranking, data transformation).
Outputs are structured dictionaries that are merged into the workflow state.
Subclasses implement an async \texttt{run()} method that returns structured data.

\subsection{Built-in Conversational Search Nodes}

Orcheo includes approximately 100 built-in nodes in total, including 45+ designed specifically for conversational search pipelines that cover ingestion, query understanding, retrieval, re-ranking, conversation management, generation, and evaluation.
Additional nodes support data transformation, external integrations (e.g., Slack, Telegram, Discord), database operations, and workflow triggers.
A comprehensive node catalog is available in the documentation at \url{https://orcheo.readthedocs.io/en/latest/node_catalog/}.

\subsection{Implementing Custom Nodes}

Researchers contribute new components by implementing custom nodes.
Listing~\ref{lst:custom-node} shows a minimal example of a custom query rewriting node:

\begin{lstlisting}[
  float,
  floatplacement=t,
  language=Python,
  basicstyle=\ttfamily\footnotesize,
  caption={A custom query rewriting node.},
  label={lst:custom-node},
  frame=single
]
from orcheo.nodes.base import AINode
from orcheo.nodes.registry import registry, NodeMetadata

@registry.register(NodeMetadata(
    name="MyQueryRewriter",
    description="Novel query rewriting method",
    category="query_processing"
))
class MyQueryRewriterNode(AINode):
    model: str = "gpt-4"
    rewrite_prompt: str = "..."

    async def run(self, state, config):
        history = state.get("messages", [])
        query = state["inputs"]["query"]
        rewritten = await self._rewrite(query, history)
        return {"rewritten_query": rewritten}
\end{lstlisting}

Key aspects of this design include:
\begin{enumerate*}[label=(\roman*)]
  \item \emph{Registry-based discovery}: The \texttt{@registry.register} decorator makes the node discoverable via the CLI without modifying core code.
  \item \emph{Pydantic configuration}: Node parameters are typed fields enabling validation and schema generation.
  \item \emph{State access}: The \texttt{run()} method receives workflow state containing conversation history, previous outputs, and inputs.
  \item \emph{Minimal dependencies}: The implementation depends only on Orcheo's base classes and standard LangChain types. A custom node typically requires only 20--50 lines of code.
\end{enumerate*}

\subsection{State Management and Data Flow}

Each Orcheo workflow uses a typed \texttt{State} schema that extends LangChain's \texttt{MessagesState}.
The schema stores workflow inputs, accumulated node results (keyed by node name), a structured response, and runtime configuration.
The \texttt{results} dictionary accumulates outputs from \texttt{TaskNode}s, keyed by node name; downstream nodes can reference upstream outputs via variable interpolation (e.g., \texttt{\{\{retriever.documents\}\}}).

\subsection{Graph Building and Execution}

Workflows are defined as directed graphs with nodes and edges.
Orcheo supports three edge types:
\begin{enumerate*}[label=(\roman*)]
  \item \emph{Sequential edges}: Simple A $\rightarrow$ B transitions.
  \item \emph{Conditional edges}: Routing based on state values (e.g., intent classification results).
  \item \emph{Parallel branches}: Concurrent execution with result aggregation.
\end{enumerate*}

The graph builder compiles workflow definitions into LangGraph's \texttt{StateGraph}, which handles execution, checkpointing, and streaming.
This architecture inherits LangGraph's benefits: async execution, automatic state persistence, and observability through LangSmith integration.
Additionally, Orcheo provides native OpenTelemetry (OTel) tracing with configurable exporters, enabling integration with observability platforms such as Jaeger or Datadog.
Canvas supports real-time trace streaming.

\subsection{Sharing Contributions}

Orcheo minimizes the code researchers need to share.
At its simplest, a researcher shares a single Python file containing a custom node; adopters can install it in their registry and immediately use it alongside built-in nodes.
Complete pipelines can also be shared as Python scripts that others import, inspect, and modify; credential references (e.g., \texttt{[[api\_key]]}) resolve at runtime from the user's vault, so workflows can be shared without exposing tokens.
The core node library (\texttt{orcheo}) can additionally be used as a standalone Python package, though the full platform is recommended for deployment and reproducibility.

\subsection{Evaluation Support}

Orcheo provides two complementary approaches for different evaluation scenarios.
Built-in evaluation nodes can be embedded directly in workflows for online monitoring, quality gating, and offline evaluation.
\texttt{ABTestingNode} routes live traffic between pipeline variants;
\texttt{LLMJudgeNode} applies configurable rubrics as real-time quality gates;
\texttt{RetrievalEvaluationNode} computes standard metrics (Recall@k, MRR, NDCG, MAP) compatible with TREC-style evaluation;
and \texttt{AnswerQualityEvaluationNode} assesses responses using reference-based metrics and faithfulness checks.
These nodes can also be composed into batch evaluation pipelines for offline benchmarking, with outputs formatted for submission to platforms like TIREx~\citep{frobe2023tirex} or ranxhub~\citep{bassani2023ranxhub}.

\heading{Reproducibility support}
Orcheo's evaluation infrastructure is designed with reproducibility in mind.
Workflow definitions, node configurations, and prompt versions are stored as versionable artifacts alongside evaluation checkpoints, enabling researchers to precisely reconstruct experimental conditions and share complete experimental setups with minimal documentation overhead.

%!TEX root = ../main.tex

\section{The Orcheo Platform}
\label{sec:platform}

Beyond the core framework, Orcheo provides a full-stack platform designed to reduce the engineering burden throughout the conversational search system lifecycle.
This section describes the platform components: backend services, the integrated ChatKit (\url{https://platform.openai.com/docs/guides/chatkit}) server and UI, SDK/CLI, deployment tools, tracing/observability capabilities, and agent skills for AI-assisted workflow development and deployment.
While the Orcheo core node library can be used independently in custom Python projects (Section~\ref{sec:framework}), the platform components described in this section provide infrastructure that makes workflows easy to deploy, share, and reproduce---capabilities that would otherwise need to be implemented and maintained by project authors.

\subsection{Architecture Overview}

\begin{figure}[t]
  \centering
  \includegraphics[width=\linewidth]{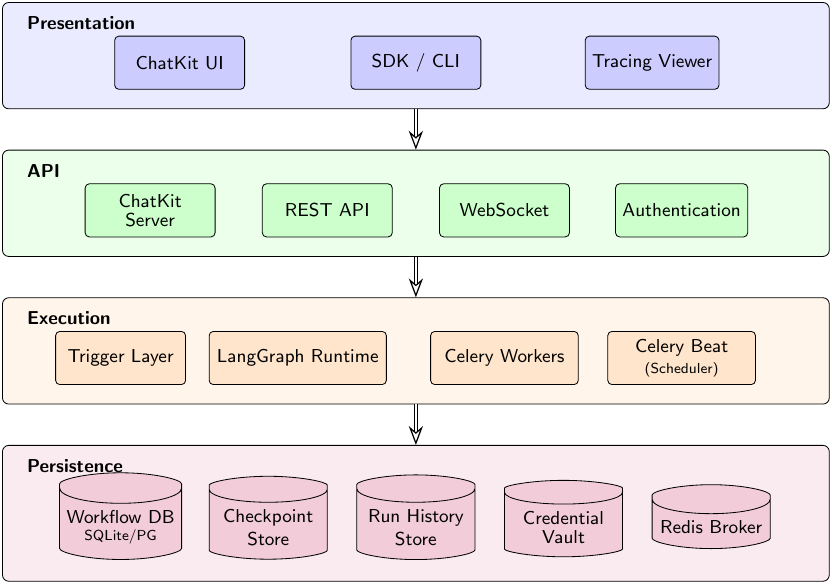}
  \vspace*{-2mm}
  \caption{Orcheo platform architecture showing the interaction between the presentation layer, backend, and execution components.}
  \Description{Layered architecture diagram with four stacked layers: Presentation (ChatKit UI, SDK/CLI, Tracing Viewer), API (ChatKit Server, REST API, WebSocket, Authentication), Execution (Trigger Layer, LangGraph Runtime, Celery Workers, Celery Beat scheduler), and Persistence (Workflow DB, Checkpoint Store, Run History Store, Credential Vault, Redis Broker), connected by arrows from top to bottom.}
  \label{fig:architecture}
\end{figure}

Figure~\ref{fig:architecture} illustrates Orcheo's platform architecture.
The system comprises four main layers:
\begin{enumerate*}[label=(\roman*)]
  \item \emph{Presentation layer}: ChatKit UI for embedding a full-featured chat experience (custom theming, widgets, tool invocation, file attachments, reasoning visualizations), as well as the SDK/CLI for workflow authoring and management and a tracing viewer for run inspection.
  \item \emph{API layer}: FastAPI services exposing REST and WebSocket endpoints, the integrated ChatKit server, and authentication.
  \item \emph{Execution layer}: Trigger layer and LangGraph runtime for orchestration, with Celery workers for background execution and Celery Beat for scheduled jobs.
  \item \emph{Persistence layer}: Workflow DB (SQLite/PostgreSQL), checkpoint store, run history store, credential vault, and Redis broker.
\end{enumerate*}

\subsection{Backend Services}

Orcheo's FastAPI backend provides:

\heading{Workflow management}
CRUD operations for workflows with versioning support.
Workflows can be created via the SDK/CLI or imported from Python scripts.

\heading{Execution engine}
Synchronous and asynchronous workflow execution with support for streaming responses.
WebSocket connections enable real-time progress monitoring and intermediate result inspection.

\heading{ChatKit server integration}
ChatKit's integration mode targets teams that need custom authentication, data residency, on-premises deployment, or bespoke orchestration.
Orcheo implements the ChatKit server contract so the ChatKit frontend can connect directly to Orcheo-hosted workflows, while the server handles request processing, tool execution, streaming responses, and persistence of threads/messages/files (including uploads and action payloads).

\heading{Tracing interface}
Each workflow run is captured as an execution trace, exposing node-level timings, inputs, and outputs.
A tracing interface surfaces these traces for debugging, provenance inspection, and performance tuning.

\heading{Background processing}
Celery workers with a Redis message broker handle long-running executions.
Celery Beat supports cron-style scheduled workflows for periodic tasks (e.g., daily index updates).

\heading{Authentication and access control}
Configurable authentication modes (disabled, optional, or required) with JWT support.
Service tokens enable programmatic access with scoped permissions.

\subsection{SDK and CLI}

The Orcheo SDK (\texttt{OrcheoClient}) provides async programmatic access for workflow execution, management, and credential handling.
The accompanying CLI (\texttt{orcheo}) exposes the same capabilities as shell commands for node discovery (e.g., \texttt{orcheo node list}), workflow management, and execution.

\subsection{AI-Assisted Workflow Development and Deployment}
\label{sec:vibe-coding}

Rather than requiring users to learn visual editor paradigms, Orcheo makes code-first development accessible through AI-assisted ``vibe-coding.''
Orcheo provides an open-source \emph{Agent Skills} repository (\url{https://github.com/AI-Colleagues/agent-skills}) that enables modern AI coding assistants (Claude Code, Cursor, OpenAI Codex CLI) to help users work with Orcheo.
With the Orcheo agent skill installed, users can issue natural language requests such as ``\emph{install Orcheo},'' ``\emph{start the full Orcheo stack with Docker},'' or ``\emph{create a conversational search workflow with query rewriting and hybrid retrieval},'' and the AI assistant generates appropriate code, commands, and configurations.
This approach offers key advantages over visual workflow tools: users describe intent in plain English rather than learning tool-specific UI conventions; generated workflows are standard Python code that can be versioned and reviewed; and the same code runs locally during development and on remote backends in production without export/import steps.

\subsection{Deployment Tools}

Orcheo includes deployment infrastructure for moving from development to production:
\begin{enumerate*}[label=(\roman*)]
\item \emph{Docker support}: Dockerfiles and Docker Compose configurations enable single-command deployment of the full stack (backend, Redis, workers, Postgres, and Canvas).
\item \emph{System units}: Production deployment templates for Linux servers with proper service management and logging.
\item \emph{Environment configuration}:
Environment variable support for database connections, authentication settings, vault configuration, and external service credentials.
\item \emph{Workflow publishing}:
Workflows can be published with configurable access controls, enabling researchers to share interactive demos through the ChatKit frontend without building bespoke frontends for each workflow.
\end{enumerate*}

\subsection{Credential Vault and Secret Management}

Conversational search systems often require access to external services (embedding APIs, vector stores, LLM providers).
Orcheo provides an AES-256 encrypted vault with runtime credential resolution via \texttt{[[credential\_name]]} references, scoped access controls, and automatic OAuth token refresh.
This separates credential management from workflow definitions, enabling secure sharing without exposing sensitive information.

\subsection{Integration Points}

Orcheo is designed to integrate with the broader conversational search research ecosystem:
\begin{enumerate*}[label=(\roman*)]
\item \emph{Vector stores}: Pluggable vector store abstraction supporting in-memory stores (development), Pinecone, and custom backends.
\item \emph{Embedding providers}: Registry-based embedding method resolution supporting LangChain embeddings, Pinecone inference, and custom implementations.
\item \emph{LLM providers}: Any LangChain-compatible LLM provider (OpenAI, Anthropic, local models via Ollama, etc.).
\item \emph{External toolkits}: The node architecture allows wrapping external IR toolkits to incorporate established components into Orcheo workflows, including built-in support for Pinecone text encoders (BM25, SPLADE).
\end{enumerate*}

%!TEX root = ../main.tex

\section{Case Studies}
\label{sec:case-studies}

We validate Orcheo's practical utility through three case studies that highlight its modularity and ease of use.
The first demonstrates how built-in nodes compose into a grounded generation pipeline; the remaining two show end-to-end evaluation on established benchmarks, illustrating how Orcheo's workflow model naturally extends to evaluation tasks while substantially reducing engineering overhead.
Companion workflows and documentation are available in the project repository and online documentation.

\subsection{Grounded Generation Pipeline}

\begin{figure}[t]
  \centering
  \includegraphics[width=\linewidth]{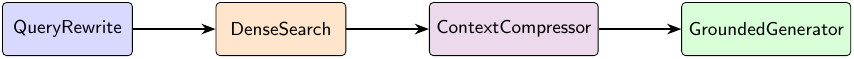}
    \vspace*{-2mm}
  \caption{Grounded generation pipeline expressed as an Orcheo graph. This pipeline is used as the system under test in the MultiDoc2Dial evaluation (Section~\ref{sec:md2d}).}
  \Description{Workflow diagram showing a linear pipeline of four nodes: QueryRewrite, DenseSearch, ContextCompressor, and GroundedGenerator.}
  \label{fig:rag-workflow}
\end{figure}

To demonstrate how Orcheo's components work together, we present a retrieval-augmented generation pipeline (Figure~\ref{fig:rag-workflow}), which is later used as the system under test in the MultiDoc2Dial evaluation (Section~\ref{sec:md2d}).

\heading{Pipeline composition}
This pipeline chains four nodes:
\begin{enumerate}[label=(\roman*), leftmargin=*]
  \item \emph{QueryRewrite} rewrites the current turn's query using conversation history.
  \item \emph{DenseSearch} retrieves passages from a vector store.
  \item \emph{ContextCompressor} deduplicates context and enforces token budgets.
  \item \emph{GroundedGenerator} produces a cited response grounded in the retrieved context.
\end{enumerate}

\heading{Modularity in practice}
Each node is independently configurable: the retrieval model, vector store, and generation model are set via a JSON configuration file, with no code changes required.
A researcher developing a novel query rewriting method can implement it as a custom \texttt{QueryRewrite} node, test it within this pipeline using Orcheo's built-in upstream and downstream nodes, and share only the custom node implementation (typically fewer than 200 lines).
Adopters swap the node into existing workflows with a one-line configuration change, demonstrating principles P1 (minimal shareable code) and P2 (composability).
Thanks to Orcheo's high-level abstractions, the entire four-stage pipeline requires only approximately 50 lines of Python code, including configuration.

\subsection{QReCC Query Rewriting Evaluation}

Our first evaluation targets query rewriting on the QReCC benchmark~\citep{anantha2021qrecc}.

\heading{Evaluation workflow}
The QReCC study is implemented as a native Orcheo workflow:
\begin{enumerate}[label=(\roman*), leftmargin=*]
  \item \texttt{QReCCDatasetNode} loads conversations with gold rewrites.
  \item \texttt{ConversationalBatchEvalNode} iterates through each conversation's turns, invokes a \texttt{QueryRewriteNode} subgraph, and collects predicted rewrites alongside gold labels.
  \item \texttt{RougeMetricsNode} and \texttt{SemanticSimilarityMetricsNode} run in \emph{parallel branches} to produce ROUGE-1 recall and embedding cosine similarity.
  \item \texttt{AnalyticsExportNode} merges results into a unified report.
\end{enumerate}
Here, ``semantic similarity'' denotes the mean cosine similarity between the embedding vectors of the predicted and gold rewrites.
This differs from the original QReCC paper, which reports Universal Sentence Encoder (USE) similarity rather than the embedding-based score used in our implementation.
The entire pipeline executes with a single \texttt{orcheo workflow run} command; the rewriting model or embedding provider can be swapped via the JSON configuration file without code changes.

\subsection{MultiDoc2Dial Grounded Generation Evaluation}
\label{sec:md2d}

Our second evaluation targets a full retrieval-generation pipeline on MultiDoc2Dial~\citep{feng2021multidoc2dial}.

\begin{figure}[t]
  \centering
  \includegraphics[width=\linewidth]{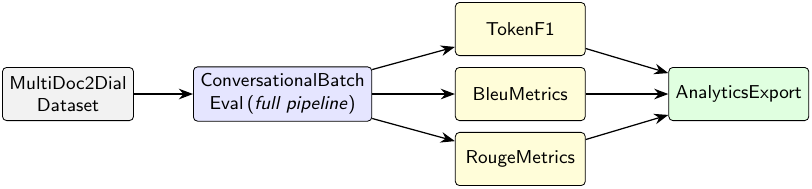}
  \vspace*{-2mm}
  \caption{MultiDoc2Dial evaluation workflow. Metric nodes execute in parallel branches and results are merged by the analytics node.}
  \Description{Evaluation workflow diagram where the MultiDoc2Dial dataset feeds into a conversational batch evaluation node, which then branches to three metric nodes (TokenF1, BleuMetrics, RougeMetrics) whose outputs are merged by an AnalyticsExport node.}
  \label{fig:eval-workflow}
\end{figure}

\heading{Nested workflow design}
This evaluation (Figure~\ref{fig:eval-workflow}) follows the same pattern but with a more complex pipeline under test: the \texttt{ConversationalBatchEvalNode} wraps the four-stage generation pipeline of Figure~\ref{fig:rag-workflow} as a composable subgraph---a workflow that evaluates another workflow.
Three metric nodes---\texttt{TokenF1}, \texttt{BleuMetrics}, and \texttt{RougeMetrics} (ROUGE-L)---run in parallel downstream, and \texttt{AnalyticsExportNode} produces the final report.
The same metric nodes used for QReCC (e.g., \texttt{RougeMetricsNode}) are reused here with different configurations, confirming their task-agnostic design.

\subsection{Evaluation Results and Engineering Efficiency}

Compared with typical manual evaluation scripts that require 500--1{,}000+ lines of code for data loading, history management, pipeline orchestration, metric computation, and reporting, equivalent Orcheo workflows require only 85--150 lines of code.
This reduction stems from composable metric nodes with a uniform output contract, automatic conversation history threading, and configuration-driven execution that separates model choices from workflow logic.

Table~\ref{tab:eval-results} reports corpus-level scores for both case studies together with published reference results.

\begin{table}[t]
\centering
\caption{Evaluation results on QReCC and MultiDoc2Dial using Orcheo workflows with GPT-4o-mini, alongside published reference results.}
\label{tab:eval-results}
\small
\begin{tabular}{@{}llcc@{}}
\toprule
\textbf{Benchmark} & \textbf{Metric} & \textbf{Orcheo} & \textbf{Published} \\
\midrule
\multirow{2}{*}{QReCC}
  & ROUGE-1 Recall            & 75.25 & 89.50 \\
  & Semantic Similarity      & 79.00 & 95.20$^\dagger$ \\
\midrule
\multirow{3}{*}{MultiDoc2Dial}
  & Token F1                & \phantom{0}8.34 & 48.04 \\
  & SacreBLEU               & 13.32 & 34.56 \\
  & ROUGE-L                 & \phantom{0}6.82 & 45.93 \\
\bottomrule
\end{tabular}

\justifying\vspace{1mm}\noindent\footnotesize
$^\dagger$QReCC published reference numbers are the Transformer++ results from the original dataset paper~\citep{anantha2021qrecc}; its semantic score is USE similarity, so this row is only an approximate point of comparison.
MultiDoc2Dial published reference numbers are the validation-seen SARCAT results~\citep{lee2024sarcat}, chosen because they report Token F1, SacreBLEU, and ROUGE-L on the same validation setting used here.
\end{table}

%!TEX root = ../main.tex

\section{Conclusion}
\label{sec:conclusion}

We have presented Orcheo, an open-source platform designed to address two persistent challenges in conversational search research: the lack of a unified framework for sharing modular contributions, and the difficulty of deploying and sharing working systems with end users.
Through its modular LangGraph-based framework with first-class AI coding support, full-stack development-to-deployment infrastructure, and a catalog of 45+ off-the-shelf CS nodes, Orcheo lowers barriers for prototyping and sharing reproducible research artifacts.
We treat conversational search pipelines as first-class workflow artifacts, and seek to accelerate research progress and facilitate knowledge transfer in conversational search.

Orcheo's current design involves deliberate trade-offs worth noting.
Performance on large-scale document collections (millions of passages) has not been extensively benchmarked.
The modular architecture promotes flexibility, but it also means researchers must make integration decisions that monolithic systems handle implicitly.
As an open-source project, Orcheo welcomes contributions to expand the ecosystem with new nodes, integrations, and best practice patterns.
Finally, while Orcheo's vibe-coding support flattens the learning curve, extending and improving the framework still requires Python proficiency.
We view these as necessary costs for the transparency and extensibility required by researchers.

We plan to extend Orcheo with:
\begin{enumerate*}[label=(\roman*)]
  \item Deeper integrations with conversational search evaluation benchmarks and shared task infrastructure, including direct TIREx submission support.
  \item A community node and workflow registry for discovering and sharing contributions.
  \item Formal node versioning with compatibility checking.% to ease upgrades.
  \item Enhanced human-in-the-loop evaluation support with annotation tools and benchmark visualization in Canvas.
\end{enumerate*}

We invite the conversational search community to adopt, extend, and contribute to Orcheo as shared infrastructure for accelerating research progress.

\subsection*{Availability}
Orcheo is released as open source at \url{https://github.com/AI-Colleagues/orcheo} under the MIT License.
The repository includes the core framework, backend services, SDK/CLI, deployment guides, and a conversational search demo suite with runnable workflows and evaluation assets (gold queries and relevance labels).

\begin{acks}
This work has been funded by the Vienna Science and Technology Fund (WWTF) under the Grant ID 10.47379/VRG24013, by the Dutch Research Council (NWO), under project numbers 024.004.022, NWA.1389.20.\-183, and KICH3.LTP.20.006, and the European Union under grant agreement No. 101201510 (UNITE). Views and opinions expressed are those of the author(s) only and do not necessarily reflect those of their respective employers, funders and/or granting authorities.
\end{acks}
%\balance
\bibliographystyle{ACM-Reference-Format}
\balance
\bibliography{orcheo}

\end{document}